\documentclass[11pt, a4paper]{amsart}
\usepackage[latin1]{inputenc}
\usepackage[all]{xy}
\usepackage{amssymb, amsmath, amscd, geometry}
\usepackage{graphicx}
\usepackage{setspace}
\usepackage{enumitem}
\usepackage{bbold}
\numberwithin{equation}{section}
\input xy
\xyoption{all}
\usepackage{latexsym}
\usepackage[usenames]{color}
\usepackage{epic} 
\usepackage{mathrsfs}
\usepackage{euscript}
\usepackage{rotating}
\usepackage{verbatim}
\newtheorem{lemma}{Lemma}[section]
\newtheorem{theorem}[lemma]{Theorem}

\newtheorem*{theorem*}{Theorem}
\theoremstyle{definition}

      \newcommand{\C}{{\mathbb C}}

\newcommand{\al}{\alpha}

\newcommand{\eps}{\epsilon}

\newcommand{\uno}{1\!\!1}

\title[Random constructions in Bell inequalities]{Random constructions in Bell inequalities: A survey}

\author{Carlos Palazuelos}

\address{Instituto de Ciencias Matem\'aticas (ICMAT)\\Departamento de An\'alisis Matem\'atico, Universidad Complutense de Madrid, 28040, Madrid, Spain}
\email{carlospalazuelos@mat.ucm.es}

\addtocounter{tocdepth}{-1}

\begin{document}

\addtolength{\parskip}{+1ex}


\maketitle

\begin{abstract}
Initially motivated by their relevance in foundations of quantum mechanics and more recently by their applications in different contexts of quantum information science, violations of Bell inequalities have been extensively studied during the last years. In particular, an important effort has been made in order to quantify such Bell violations. Probabilistic techniques have been heavily used in this context with two different purposes. First, to quantify how common the phenomenon of Bell violations is; and secondly, to find large Bell violations in order to better understand the possibilities and limitations of this phenomenon. However, the strong mathematical content of these results has discouraged some of the potentially interested readers. The aim of the present work is to review some of the recent results in this direction by focusing on the main ideas and removing most of the technical details, to make the previous study more accessible to a wide audience.
\end{abstract}
\section*{Introduction}
Bell inequalities have attracted much attention in the last years. Their original interest as a key tool in the study of foundations of quantum mechanics has been nowadays surpassed by the relevance of these inequalities in different contexts such us quantum cryptography, communication complexity protocols and generation of trusted random numbers. In addition, Bell inequalities have been shown to be intimately related with some problems in computer science and operator algebras theory, capturing in this way the interest from those communities. Given the great importance of probabilistic techniques in those fields as well as in different areas of quantum information, it is not surprising that they are also very useful in the context of Bell inequalities. In fact, random constructions have been a key tool to solve some questions which had remained open for a long time in the field. However, despite their potential usefulness, these techniques and results are still far from being considered natural by many people working on quantum nonlocality. The aim of the present work is to review some of the most important results in the context of Bell inequalities for which probabilistic techniques have played a crucial role. Here, we will focus on the main ideas without paying attention to the technical details with the hope that this makes the previous works more appealing for the non-experts. Hence, this work must not be understood as a general survey on Bell inequalities, for which the reader can find excellent references in \cite{BCPSW}, \cite{BCMW}, \cite{Tsirelson}, \cite{WeWo I}. In particular, we will deliberately skip some standard topics such as connections with other areas, physical applications of the results and so on, with the upside of going directly to the important points of our discussion.

Let us start by introducing the basic objects of study. Bell inequalities  were first considered by Bell in \cite{Bell} as a way to clarify an apparently metaphysical dispute on the completeness of quantum mechanics as a model of Nature arising from the work \cite{EPR}. Given two spatially separated quantum systems, controlled by Alice and Bob, respectively, and described by a bipartite quantum state $\rho$, Bell showed that certain probability distributions  obtained from an experiment in which Alice and Bob perform some measurements $x$ and $y$ in their corresponding systems with possible outputs $a$ and $b$, respectively, cannot be explained by a classical model\footnote{Formally, Bell talked about a local hidden variable model (LHVM).}. More precisely, if $P=(P(a,b|x,y))_{x,y; a,b=1}^{N,K}$ denotes the probability distribution\footnote{Note that $P$ is not a probability distribution itself. For every $x,y$ fixed we have that $(P(a,b|x,y))_{a,b=1}^K$ is a probability distribution. However, it is standard to use this terminology.} obtained in such an experiment, we say that $P$ is a \emph{classical (or local) probability distribution} if it can be written as
\begin{equation*}\label{classical}
P(a,b|x,y)=\int_\Omega P_\omega(a|x)Q_\omega(b|y)d\mathbb{P}(\omega)
\end{equation*}
for every $x,y,a,b$, where $(\Omega, \mathbb{P})$ is a probability space and for every $\omega\in \Omega$ we have $P_\omega(a|x)\ge 0$ and $\sum_a P_\omega(a|x)=1$ for all $a,x$ (resp. $Q_\omega(b|y)\ge 0$ and $\sum_b Q_\omega(b|y)=1$ for all $b,y$). We denote the set of classical probability distributions by $\mathcal{L}$. It is easy to see that $\mathcal{L}$ is a polytope (convex set with a finite number of extreme points) and the inequalities describing its facets are called \emph{Bell inequalities}. On the other hand, we say that $P$ is a \emph{quantum probability distribution} if
\begin{equation*}
P(a,b|x,y)=tr(E_x^a\otimes F_y^b \rho)
\end{equation*}
for every $x,y,a,b$, where  $\rho$ is  a density operator acting on the tensor product of two Hilbert spaces $H_1\otimes H_2$ and $(E_x^a)_{x,a}$, $(F_y^b)_{y,b}$ are two sets of operators representing POVM measurements acting on $H_1$ and $H_2$ respectively. That is, $E_x^a\geq 0$ and $\sum_{a}E_x^a=\uno_{H_1}$ for all $a, x$ (resp. $F_y^b\geq 0$ and
$\sum _{b}F_y^b=\uno_{H_2}$ for all $b,y$). We denote the set of quantum probability distributions by $\mathcal{Q}$. It is easy to see that this set is convex and it verifies $\mathcal L\subset \mathcal{Q}$. As it was shown by Bell, the converse inclusion fails; equivalently, there exist quantum probability distributions violating some Bell inequalities. 

In fact, in the beginning of this theory a slightly simpler scenario was considered. In the particular case where Alice's and Bob's measurements are binary (that is, $a_x,b_y=\pm 1$ for every $x,y$) one can consider the joint correlation:
\begin{align*}
\gamma_{x,y}=\mathbb E[a_x\cdot b_y]=P(1,1|x,y)+P(-1,-1|x,y)-P(-1,1|x,y)+P(1,-1|x,y)
\end{align*}for every $x,y=1,\dots, N$. By plugging the definition of classical (resp. quantum) probability distribution in the previous expression one justifies the definition of \emph{classical (or local) correlation matrices} as those which can be expressed in the form
\begin{align*}
\gamma_{x,y}=\int_\Omega A_x(\omega)B_y(\omega)d\mathbb{P}(\omega)
\end{align*}
for every $x,y$, where $(\Omega, \mathbb{P})$ is a probability space and for every $\omega\in \Omega$ we have $A_x(\omega)\in \{-1,1\}$ for every $x$ (resp. $B_y(\omega)\in \{-1,1\}$ for every $y$); and \emph{quantum correlation matrices} as those of the form
\begin{align*}
\gamma_{x,y}=tr(A_x\otimes B_y\rho), \text{     }
\end{align*}for every $x,y$, where  $\rho$ is  a density operator acting on the tensor product of two Hilbert spaces $H_1\otimes H_2$ and $(A_x)_x$, $(B_y)_y$ are self-adjoint operators acting on $H_1$ and $H_2$ respectively verifying $\max_{x,y}\{\|A_x\|, \|B_y\|\}\leq 1$. Since the different contexts will be clear along the work, we will also denote by $\mathcal L$ and $\mathcal Q$ the set of classical and quantum correlation matrices respectively. Note that in order to consider this simpler context we made two simplifications: We restrict to binary measurements and we only considered the joint correlations of Alice's and Bob's measurements (and not the marginals). In particular, $\mathcal L$ is again a polytope defined by its facets, now called \emph{correlation Bell inequalities}, contained in the convex set $\mathcal Q$. Bell's work actually showed that this inclusion is strict even for the simplest case $x,y=1,2$ (which implies the result for probability distributions).

In this work it will be useful to understand Bell inequalities in a slightly more general sense as \emph{dual objects} of probability distributions. In particular, \emph{any} real tensor $M=(M_{x,y}^{a,b})_{x,y;a,b=1}^{N,K}$ (resp. real matrix $M=(M_{x,y})_{x,y=1}^{N}$) defines a Bell inequality (resp. correlation Bell inequality) by considering the dual action on probability distributions (resp. correlation matrices): $\langle M,P\rangle=\sum_{x,y;a,b}M_{x,y}^{a,b}P(a,b|x,y)$ (resp. $\langle M,\gamma\rangle=\sum_{x,y} M_{x,y}\gamma_{x,y}$). Given $M$, we will denote its \emph{classical value} and its \emph{quantum value} respectively by
\begin{align*}
\omega(M)=\sup\big\{|\langle M,Z\rangle|: Z\in \mathcal L\big\}\text{     }\text{     }\text{   and     }\text{     } \text{     }\omega^*(M)=\sup\big\{|\langle M,Z\rangle|: Z\in \mathcal Q\big\}.
\end{align*}
This picture allows us not only to describe Bell inequality violations: $$M, \text{     } LV(M):=\frac{\omega^*(M)}{\omega(M)}>1,$$but also to quantify these violations by means of the above quantity $LV$. As we will explain below, this quantification is crucial if one wants to understand the possibilities and limitations of Bell inequalities. The most famous correlation Bell inequality, the \emph{CHSH inequality} \cite{CHSH}, is given by the $2\times 2$ matrix $M$ defined by $M_{2,2}=-1$ and $M_{x,y}=1$ otherwise, for which one can check that $\omega(M)=2$ and $\omega^*(M)=2\sqrt{2}$. Thus, $LV(M)=\sqrt{2}$. 

Although the previous descriptions only consider the bipartite case, we can extend all the previous definitions to the multipartite setting straightforward. The multipartite scenario will be very important in our work because of two reasons. First of all, even in the tripartite case we will observe new phenomena which cannot be found in the scenario of two parties. Secondly, considering $n$ parties introduces a new parameter in the problem and, as we will see, the answer for different questions can strongly depend on it.

Once we know that violations of Bell inequalities exist, it is natural to wonder how common this phenomenon is. As we explained previously, there are two key objects in our picture, Bell inequalities (resp. correlation Bell inequalities) and probability distributions (resp. correlation matrices), which lead to different questions. On the one hand, one can study the probability of having $LV(M)>1$ if we pick a Bell inequality (resp. correlation Bell inequality) $M$ at random or, even more, which value $LV(M)$ we should expect if we follow this procedure. On the other hand, one can pick a quantum probability distribution (resp. quantum correlation matrix) $Q$ at random and ask how likely the event $Q\notin \mathcal L$ is. However, two main obstacles appear when studying these questions. 

The first problem is that it is not so clear what picking these objects at random means. For that, one has to consider a probability measure on the corresponding sets and that is not always easy nor natural. In fact, the sampling procedure is particularly tricky for quantum probability distributions (resp. quantum correlation matrices) since here one has to impose a certain structure which is not required when one samples Bell inequalities. Regarding the definition of quantum probability distributions and quantum correlation matrices above one could think about sampling states and measurements independently. However, this procedure presents several problems such as deciding the dimension of the corresponding Hilbert spaces, the lack of a natural way to sample general measurements and so on. Some works have also considered the problem of fixing one of these objects (either the state or the measurements) and sampling the other one at random. In fact, considering random measurement seems to be an interesting problem from the experimental point of view since it is directly connected to the absence of a shared reference frame in Bell experiments.

A second problem is that the quantities $\omega(M)$, $\omega^*(M)$ and $LV(M)$ are in general very difficult to compute. The study \cite{Tsirelson} performed by Tsirelson clarified the context of bipartite correlation Bell inequalities. In particular, he gave an alternative description for the quantity $\omega^*(M)$ which allows to understand the quantity $LV_2(N)=\sup\big\{LV(M): M=(M_{x,y})_{x,y=1}^{N}\big\}$\footnote{The subscript $2$ in $LV_2(N)$ denotes that it is in the bipartite case.} via the so called \emph{Grothendieck constant}. However, the situation is more intricate in the case of general bipartite Bell inequalities as well as in the tripartite correlation case. If we define $LV_2(N,K)=\sup\big\{LV(M): M=(M_{x,y}^{a,b})_{x,y;a,b=1}^{N,K}\big\}$, it is not clear how large this quantity can be as a function of $N$ and $K$ and the same happens for the quantity $LV_3(N)$ in the tripartite correlation scenario. Hence, some questions \emph{should} be answered before trying to understand the probabilistic behavior of Bell violations in these contexts. Interestingly, probabilistic techniques are also very important for this purpose since they have been shown to be a useful tool to study how large the quantities $LV_3(N)$ and $LV_2(N,K)$ can be. On the other hand, beyond its importance from the previous point of view, looking for large Bell violations is an interesting question itself, since the quantity $LV$ can be understood as a quantification of how better one can solve certain tasks if quantum resources are used instead of classical ones (see \cite{JP}, \cite{JPPVWII}, \cite{PWPVJ} for more information).

Finally, let us remark that, despite the restriction imposed in the title of this survey, the number of works on Bell inequalities using probabilistic tools is very large. Here, we will focus on those results studying asymptotic behaviors. That is, those results describing the probabilistic nature of a problem when (some of) the parameters go to infinity.  With this in mind, let us remind the reader the standard asymptotic notation, which will be constantly used in this work. Given two nonnegative functions $f(n)$ and $g(n)$ on natural numbers, we will write $f(n)=O\big(g(n)\big)$ (resp. $f(n)=\Omega \big(g(n)\big)$) if there exist a constant $C$ and a natural number $n_0$ such that $f(n)\leq Cg(n)$ (resp. $f(n)\geq Cg(n)$) for every $n\geq n_0$. On the other hand, we will write $f(n)=o\big(g(n)\big)$ if $\lim_{n\rightarrow \infty}f(n)/g(n)=0$.

The survey is organized as follows. Section \ref{Section: Bipartite correlation Bell inequalities} is devoted to explaining some results about the probabilistic nature of quantum nonlocality in the bipartite correlation case. As we will see, the deep understanding of this situation thanks to Tsirelson's work allows to study the problem in a very natural way. In Section \ref{multipartite correlation Bell inequalities} some results about lower an upper bounds for tripartite correlation Bell inequality violations will be reviewed. This will lead us to the first examples of \emph{unbounded violations of tripartite correlation Bell inequalities}. Section \ref{Section: Correlation Bell inequalities for a large number of parties} will deal with some results about the probabilistic nature of multipartite Bell inequalities in some particular cases as a function of the number of parties $n$. Finally, in Section \ref{General bipartite Bell inequalities} some recent results about general bipartite Bell inequalities and the quantity $LV_2(N,K)$ will be reviewed. In particular, we will explain some random constructions in this setting and compare them with some new results from computer science.  
\section{Bipartite correlation Bell inequalities}\label{Section: Bipartite correlation Bell inequalities}
As we mentioned in the Introduction Bell's work showed the strict inclusion $\mathcal L\varsubsetneq \mathcal Q$ (see also \cite{CHSH}). This picture was completed by Tsirelson in the case of correlation matrices, by showing that the set $\mathcal Q$ is not much bigger than the set $\mathcal L$. More precisely, one has $\mathcal L\varsubsetneq \mathcal Q\varsubsetneq K_G^\mathbb R\mathcal L,$ where $1.67696...\leq K_G^\mathbb R \leq 1.78221...$ is the real Grothendieck constant\footnote{The exact value of the Grothendieck constant is still unknown in both the real and the complex case (see \cite{BMMN} for the most recent progress).}. The following result is the standard statement of the corresponding theorem.
\begin{theorem}[Tsirelson]\label{Tsirelson-Gro}
\begin{align}
LV_2(N):=\sup\Big\{\frac{\omega^*(M)}{\omega(M)}: M=(M_{x,y})_{x,y=1}^N\Big\}\leq K_G^\mathbb R  \text{    } \text{    for every   } \text{    } N.
\end{align}
\end{theorem}In fact, the Grothedieck constant can be defined by $K_G^\mathbb R:=\sup_N LV_2(N)$. Theorem \ref{Tsirelson-Gro} is a consequence of Grothendieck's inequality and a result proved by Tsirelson \cite{Tsirelson} which states that $\gamma=(\gamma_{x,y})_{x,y=1}^N$ is a quantum correlation matrix if and only if there exist a real Hilbert space $H$ and unit vectors $u_1,\cdots, u_N, v_1,\cdots, v_N$ in $H$ such that
\begin{align}\label{Tsirelson description}
\gamma_{x,y}=\langle u_x,v_y\rangle \text{       } \text{    for every    }x,y=1,\cdots, N.
\end{align}
In  \cite{Tsirelson} the author posed the open question of whether a similar result to Theorem \ref{Tsirelson-Gro} holds in the tripartite case. This is related to the lack of Grothendieck's inequality for trilinear forms and we will go over this question in the following section.

In \cite{Ambainis I} the authors tackled the question of how likely it is for a random bipartite correlation Bell inequality\footnote{The work \cite{Ambainis I} deals with bipartite XOR games, but the problem is completely equivalent.} $M$ to verify that the quotient between $\omega^*(M)$ and $\omega(M)$ is strictly larger than one. In order to study this problem, one first needs to define a way of sampling these inequalities. In \cite{Ambainis I}, the authors considered random $N\times N$ matrices $M$ sampled from $\{-1,1\}^{N^2}$ with respect to the uniform measure. In fact, although the authors focused on sign matrices, the same techniques can be applied to study more general random matrices, such as real gaussian matrices. The main result in \cite{Ambainis I} states as follows.
\begin{theorem}\label{theorem almost all XOR}
If $M=\big(\epsilon_{x,y}\big)_{x,y=1}^N$ is a random matrix sampled from $\{-1,1\}^{N^2}$ with respect to the uniform measure, then
\begin{align*}
\lim_{N\rightarrow \infty}\mathbb P\Big\{M:1.5638...>\frac{\omega^*(M)}{\omega(M)}>1.2011...\Big\}=1.
\end{align*}
\end{theorem}

Theorem \ref{theorem almost all XOR} implies that for almost any correlation Bell inequality its quantum value is strictly large than its classical one. On the other hand, according to Theorem \ref{Tsirelson-Gro} and the comments below, the previous equation also tells us that with probability tending to one these inequalities are \emph{not so close} to the optimal value $K_G^\mathbb R\in [1.67, 1.79]$ for the quotient between $\omega^*(M)$ and $\omega(M)$. Let us also mention that, although very natural, the way of sampling bipartite correlation Bell inequalities considered in \cite{Ambainis I} is very particular. Indeed, this procedure does not consider all bipartite correlation Bell inequalities since, for instance, all the entries of the matrices have the same absolute value $1$. One can think about some other ways of sampling them by considering for instance a bipartite correlation Bell inequality $M\in \mathbb R^{N^2}$ as an element in the unit sphere with respect to the norm $\|M\|=\sum_{x,y=1}^N|M_{x,y}|$. Then, one can naturally define a probability distribution on this sphere. Preliminary calculations suggest that similar results to Theorem \ref{theorem almost all XOR} could be obtained in that context.

Since the most important part of Theorem \ref{theorem almost all XOR} is the lower bound for the quotient, we will just briefly explain how to prove that estimate. To this end, the authors first show that $\lim_{N\rightarrow \infty}\mathbb P\big\{M:\omega(M)\leq \big(1.6651...+o(1)\big)N^{\frac{3}{2}}\big\}=1$. This is obtained as an application of the Chernoff bound to the random variable $\sum_{x,y=1}^N\epsilon_{x,y}t_xs_y$ for a fixed choice of signs $t_x, s_y=\pm 1$, $x,y=1,\cdots ,N$; and a counting argument  to consider the $2^{2N}$ possible choices of signs. In order to prove the estimate $\lim_{N\rightarrow \infty}\mathbb P\big\{M:\omega^*(M)\geq \big(2-o(1)\big)N^{\frac{3}{2}}\big\}=1$, from where one obtains the lower bound in Theorem \ref{theorem almost all XOR}, the authors used a clever construction based on the Marcenko-Pastur law, which describes the behavior of the singular values of the random matrix $M$. If we call $L$ and $R$ the $n\times m$ matrices whose columns are respectively the left and right singular vectors of the matrix $M$ associated to its $m$ largest singular values, basic linear algebra shows that
\begin{align}\label{Ambainis trick}
\sum_{x,y=1}^N M_{x,y}\langle u_x, v_y\rangle=\sum_{i=1}^m\lambda_i,
\end{align}where the $m$-dimensional vectors $u_x$'s and $v_y$'s are the rows of $L$ and $R$ respectively and $(\lambda_i)_{i=1}^m$ are the corresponding singular values. On the other hand, the Marcenko-Pastur law \cite{MaPa} tells us that if one defines the function $f(s)=\frac{1}{2\pi}\int_{s^2}^4\sqrt{\frac{4}{x}-1}dx$ on $[0,2]$, for every $\epsilon>0$ the number $m$ of singular values verifying $\lambda_i>(2-\epsilon)\sqrt{N}$ belongs to the interval $\big[\big( f(2-\epsilon)-o(1)\big)N,\big( f(2-\epsilon)+o(1)\big)N\big]$ with probability tending to 1 as $N$ goes to infinity. Hence, the quantity (\ref{Ambainis trick}) is lower bounded by $(2-\epsilon)\big( f(2-\epsilon)-o(1)\big)N^{\frac{3}{2}}$ with probability tending to one. The technical part of the proof in \cite{Ambainis I} consists of adapting the Marcenko-Pastur law to show that one can assume that for a given $\delta>0$, all the previous vectors $u_x$'s and $v_y$'s have norm lower than or equal to $\sqrt{f(2-\epsilon)+\delta}$ with probability tending to one\footnote{In fact, the modified Marcenko-Pastur law proved in \cite[Theorem 3]{Ambainis I} gives the probability for each of these vectors to have norm larger than $ \sqrt{f(2-\epsilon)+\delta}$. Then, the authors proved that Eq. (\ref{Ambainis trick}) is not affected if one rules out those vectors with large norm.}. Therefore, by considering the normalized version of the previous vectors  $\tilde{u}_x$, $\tilde{v}_y$, one obtains that the quantum correlation $\gamma=\big(\langle\tilde{u}_x,\tilde{v}_y\rangle\big)_{x,y=1}^N$ verifies 
\begin{align*}
\omega^*(M)\geq \sum_{x,y=1}^N M_{x,y}\gamma_{x,y}=\frac{1}{f(2-\epsilon)+\delta}\sum_{i=1}^m\lambda_i\geq \frac{(2-\epsilon)\big( f(2-\epsilon)-o(1)\big)}{f(2-\epsilon)+\delta}N^{\frac{3}{2}}.
\end{align*}Since $\epsilon$ and $\delta$ can be made arbitrarily small one obtains the desired estimate.

Interestingly, the authors show in \cite{Ambainis I} that $(2+o(1))N^{\frac{3}{2}}$ is also an upper bound for the quantum value $\omega^*(M)$, so it is optimal. However, the best lower bound for the classical value $\omega(M)$ obtained in \cite{Ambainis I} is far from the upper bound explained above (see \cite[Section 4]{Ambainis I} for details). Obtaining the exact asymptotic value of $\omega(M)$ looks a difficult open problem, which seems more interesting from a mathematical and computer scientific point of view than from a physical perspective.

In \cite{GJPV} the same problem was considered from the correlation matrices point of view. That is, if one picks a quantum correlation matrix at random, which is the probability that it is nonlocal? The equivalent reformulation (\ref{Tsirelson description}) of a quantum correlation gives a natural sampling procedure which avoids most of the problems that we mentioned in the Introduction: one can pick the  vectors $u_1,\cdots, u_N, v_1,\cdots, v_N$ independently and uniformly distributed on the unit sphere of $\mathbb R^m$. It is well known that this is exactly the same as sampling independent normalized $m$-dimensional real gaussian vectors. It is very easy to see that if one fixes any finite $m$, the probability that a quantum correlation matrix sampled according to the previous procedure is nonlocal tends to one as $N$ tends to infinity (see \cite[Section 2]{GJPV} for details). However, this kind of sampling does not say too much since the set of quantum correlation matrices of order $N$ which can be obtained with a fixed $m$ is very small. The interesting case is that where $m$ and $N$ are of the same order. The main result in \cite{GJPV} gives indeed and answer to the considered problem as a function of $\alpha=\frac{m}{N}$: 
\begin{theorem}\label{Main result quantum correlations}
Let $u_1,\cdots, u_N, v_1,\cdots, v_N$ be  $2N$ vectors sampled independently according to the uniform measure on the unit sphere of $\mathbb R^m$ and denote by $\gamma=(\langle u_i,v_j\rangle)_{i,j=1}^N$ the corresponding quantum correlation matrix. Then, if we denote $\alpha=\frac{m}{N}$ we have
\begin{itemize}[leftmargin=*]
\item[a)] If $\alpha\leq \alpha_0\approx 0.004$, then $\gamma$ is nonlocal with probability tending to one as $N$ tends to infinity.
\item[b)] If $\alpha> 2$, then $\gamma$ is local with probability tending to one as $N$ tends to infinity.
\end{itemize}
\end{theorem}

There is a considerable gap between $\alpha_0$ and $2$. In fact, one should not expect this result to be optimal. This is because the proof of part a) above is based on an approximation of gaussian matrices by orthogonal ones and a use of the main result in \cite{Ambainis I}, where $N$ must be \emph{artificially} larger than $m$. However, the important point of the previous statement is that it shows a nontrivial phase transition for the nonlocal properties of $\gamma$ as a function of the parameter $\alpha=\frac{m}{N}$. Clarifying the case $\alpha=1$, studying any possible relation between $\alpha$ and $K_G^\mathbb R$ and reducing the gap appearing in Theorem \ref{Main result quantum correlations} are proposed in \cite{GJPV} as open questions.

In order to understand how part a) above can be proved, let us go back to Theorem \ref{theorem almost all XOR} and to reformulate it in a different way. Let us assume that we sample two independent $N\times N$ orthogonal matrices $U$ and $V$ according to the Haar measure. Then, if we call $u_i$ (resp. $v_j$) the normalized vector formed by the first $m$ coordinates of the $i$-th row of $U$ (resp. $j$-th row of $V$), then the quantum correlation $\gamma=\big(\langle u_i, v_j\rangle\big)_{i,j=1}^N$ is not local with probability tending to 1 as $N$ goes to infinity. Here, $m$ is the one considered in Eq. (\ref{Ambainis trick}) above. Indeed, this can be deduced from the fact that if we consider the singular value decomposition of a real random gaussian matrix $G=UDV$ (for which Theorem \ref{theorem almost all XOR} can be stated exactly in the same way), the matrix $U$ and $V$ are independent Haar distributed orthogonal matrices (see \cite[Propostion 2.1]{GJPV} for details). However, this fact cannot be used directly in \cite{GJPV}, since in the problem considered there one must start sampling gaussian matrices $G_1$ and $G_2$ instead of orthogonal matrices $U$ and $V$. More precisely, according to the comments above, the problem considered in \cite{GJPV} consists of studying the nonlocal properties of the correlation $\big(\langle g_i, h_j\rangle\big)_{i,j=1}^N$, where $g_i$ (resp. $h_j$) are the normalized vectors formed by the $m$ first coordinates of the $i$-th row of an $N\times N$ gaussian matrix $G$ (resp. $j$-th row of an $N\times N$ gaussian matrix $H$). However, part a) above can be obtained by invoking \cite[Theorem 1.1]{GPV}, which shows that there is a \emph{coupling} between real gaussian matrices $G$ and Haar distributed orthogonal matrices $O$ such that the norm $\sup_{i=1,\cdots, N}\big\|F_i^m(G-\sqrt{N}O)\big\|$ is controlled, where here $F_i^m(G-\sqrt{N}O)$ is the $i$-$th$ row of the matrix $G-\sqrt{N}O$ truncated to its first $m$ entries. Indeed, a suitable control on the previous norm allows the authors in \cite{GJPV} to replace the nonlocal correlation $\gamma$ introduced above from two orthogonal matrices by another nonlocal correlation $\tilde{\gamma}=\big(\langle g_i, h_j\rangle\big)_{i,j=1}^N$, where the $g_i$'s and $h_j$'s are now $m$-dimensional normalized real gaussian vectors (see \cite[Theorem 2.3]{GJPV} for details).

Part b) of Theorem \ref{Main result quantum correlations} can be obtained by using classical Banach space techniques.
\section{Tripartite correlation Bell inequalities: Unbounded violations}\label{multipartite correlation Bell inequalities}
\subsection{Extensions of Tsirelson's result to the multipartite setting}
In order to study Bell violations in the multipartite case and regarding the importance of the GHZ state in the setting of two parties, it seems very reasonable to consider the generalized $n$-partite $d$-dimensional GHZ state: $|\psi\rangle=\frac{1}{\sqrt{d}}\sum_{i=1}^d|i\rangle^{\otimes n}$. On the other hand, any possible strategy looking for unbounded Bell violations should definitely exploit the absence of \emph{the} Grothendieck inequality in the multilinear framework. However, one should precise this statement a little bit. Grothendieck's inequality can be stated in many equivalent ways (see \cite[Page 172]{DeFl}) and it turns out that, in the multilinear case, the corresponding generalizations of those statements are not equivalent anymore. In fact, although several of the possible extensions of Grothendieck's inequality to the multilinear setting have been proved to be false, \emph{a few of them} remain valid in the new context (see \cite{Blei}, \cite{BPV}, \cite{Carne}, \cite{Tonge}). The following generalization of Grothendieck's inequality was proved in \cite{Blei}, \cite{Tonge}\footnote{In fact, we state here a slightly modified version for $T$ real, which involves a modification in the constant (see \cite[Theorem 9]{BBLV} for details).}.
\begin{theorem}\label{Gro Mult}
Let $n, N\geq 2$ and $d$ be positive integers, let $T=(T_{i_1,\cdots,i_n})_{i_1,\cdots,i_n=1}^N$ be a real tensor and $x_{i_1},\cdots ,x_{i_n}$ elements in the unit ball of a complex Hilbert space $H$ of dimension $d$ for every $i_1,\cdots,i_n=1,\cdots, N$. Then,
\begin{align*}
\Big|\sum_{i_1,\cdots, i_n=1}^NT_{i_1,\cdots,i_n}\langle x_{i_1},\cdots , x_{i_n}\rangle\Big|\leq 2^{\frac{3n-5}{2}}K_G^{\mathbb C} \cdot \omega (T),
\end{align*}where $\langle x_{i_1},\cdots , x_{i_n}\rangle=\sum_{k=1}^dx_{i_1}(k)\cdots x_{i_n}(k)$ and $K_G^{\mathbb C}$ is the complex Grothendieck constant.
\end{theorem}

In \cite[Theorem 11]{PWPVJ} the authors used another version of Theorem \ref{Gro Mult} to show that the largest Bell violation achievable by three parties\footnote{The result can be generalized to $n$ parties straightforward.} sharing a $d$-dimensional GHZ state is upper bounded by $4\sqrt{2}K_G^{\mathbb C}$, independently of the number of inputs $N$ and the dimension $d$. This was the first result proving that a nontrivial family of states can not give large Bell violations and it can be seen as a generalization of Theorem \ref{Tsirelson-Gro}. In \cite{PWPVJ} the authors posed the problem of whether an analogous result could be proved for Schmidt states: $|\psi_\alpha\rangle=\sum_{i=1}^d\alpha_i|i\rangle^{\otimes n}$, where $\alpha=(\alpha_i)_{i=1}^d$ verifies $\sum_{i=1}^n|\alpha_i|=1$; and provided in addition a direct connection between such a problem and an open question in the context of operator algebras. Two years latter this question was answered in \cite{BBLV} by using a surprisingly easy argument. The authors in \cite{BBLV} realized that the upper bound for the GHZ state can be obtained in a straightforward manner from Theorem \ref{Gro Mult} and extended the result to Schmidt states by using a nice expansion of each of these states in terms of non-normalized GHZ states (see \cite[Theorem 1]{BBLV}).
\begin{theorem}\label{Schmidt bounded}
Let $T$ be an $n$-partite correlation Bell inequality. Then, for every $n$-partite quantum correlation $\gamma$ constructed with the state $|\psi_\alpha\rangle=\sum_{i=1}^d\alpha_i|i\rangle^{\otimes n}$ we have that $\langle T, \gamma\rangle\leq 2^{\frac{3n-5}{2}}K_G^{\mathbb C}\cdot \omega(T)$, independently of the number of inputs $N$ and the local dimension $d$.
\end{theorem}
In \cite{BBLV} an exhaustive study of Carne's extension of the Grothendieck inequality to the multilinear case \cite{Carne} was performed to conclude that an analogous result to Theorem \ref{Schmidt bounded} can also be stated when the parties share a \emph{clique-wise entangled state} (see \cite[Theorem 2]{BBLV}). This implies in particular that for tripartite correlation Bell inequalities, the amount of Bell violation achievable by an arbitrary stabilizer states is uniformly bounded. 
\subsection{Unbounded violations of tripartite Bell inequalities}\label{section Unbounded violation of tripartite Bell inequalities}
The previous results rule out most of the candidates one would first study in order to find large Bell violations in the multipartite setting. Then, as in many other contexts, it becomes natural to study the behavior of \emph{random states}. A standard way of sampling random pure quantum states $|\psi\rangle \in \mathbb C^d$ is by using the uniform measure on the unit sphere of the Hilbert space $\mathbb C^d$ or, equivalently\footnote{One needs to normalize the gaussian state to have the same distribution.}, using gaussian random variables $|\psi\rangle=\sum_{i=1}^dg_i |i\rangle\in \mathbb C^d$. When we are dealing with more than one system this last notation is usually extended to $|\psi \rangle=\sum_{i_1,\cdots, i_n=1}^dg_{i_1,\cdots, i_n} |i_1\cdots i_n\rangle\in \mathbb (\mathbb C^{d})^{\otimes_n}$. In the case of two and three systems it is also very common to use random unitaries sampled according to the Haar measure in the unitary grupo $\mathbb U_d$. We write $|\psi\rangle=\sum_{i,j=1}^dU(i,j) |ij\rangle\in \mathbb (\mathbb C^{d})^{\otimes_2}$ or $|\psi\rangle=\sum_{i,j,k=1}^dU_i(j,k) |ijk\rangle\in \mathbb (\mathbb C^{d})^{\otimes_3}$, where in the last expression the unitaries $(U_i)_{i=1}^d$ are sampled in the cartesian product $\prod_d \mathbb U_d$. The unitary approach has been extensively used in quantum channel theory and entanglement theory.

In \cite{PWPVJ} it was proved for the first time that, in contrast to the bipartite case, there are tripartite correlation Bell inequalities which lead to unbounded violations. 
\begin{theorem}\label{Unbounded Bell violation}
For every dimension $d\in \mathbb N$, there exist $D \in  \mathbb N$, a pure state $|\psi\rangle\in \mathbb C^{d}\otimes \mathbb C^D\otimes \mathbb C^D$ and a Bell inequality $T=(T_{i,j,k})_{i,j,k=1}^{2^{d^2}, 2^{D^2}, 2^{D^2}}$ such that the violation by $|\psi\rangle$ on such an inequality is $\Omega\big(\sqrt{d}\big)$.
\end{theorem}
This theorem implies that there does not exist a uniform constant $C$ such that $LV_3(N)\leq C$ independently of the number of inputs $N$; giving in this way a negative answer to the question posed by Tsirelson in \cite{Tsirelson}. It is advisable to extend the previous definition to $LV_3(N, d)=\sup\big\{\frac{\omega_{d}^*(T)}{\omega(T)}: T=(T_{x,y,z})_{x,y,z=1}^{N}\big\}$, where here $\omega_{d}^*(T)$ denotes the quantum value of $T$ when we only consider tripartite quantum correlations constructed with quantum state
of local dimension $d$. Then, it is important to point out that, in order to obtain unbounded violations, one must allow to increase both the number of inputs $N$ and the dimension of the Hilbert space of the system $d$. Indeed, if we fix one of these parameters, then the amount of violation is upper bounded by a constant depending on it.
\begin{theorem}\label{upper bounds tripartite} 
The following upper bound holds:
$$LV_3(N, d)=O\big(\sqrt{k}\big), \text{      }    \text{  where  } \text{      }   k=\min\big\{N, d\big\}.$$
\end{theorem}
The upper bound as a function of $d$ was first proved in  \cite{PWPVJ} (see \cite[Theorem 3]{BrVi} for an alternative proof). In fact, one can state a stronger result since it suffices that only one party has local dimension $d$. This tells us that Theorem \ref{Unbounded Bell violation} is optimal in the local dimension $d$. The (easier) upper bound as a function of $N$ can be found in \cite[Theorem 2]{BrVi} and it also admits an extension requiring that only one party has $N$ inputs. In \cite{BrVi}  the authors generalized the previous theorem to $n$ parties by providing the upper bound $O(d^\frac{n-2}{2})$ whenever the share state of the parties is restricted to have local dimension $d$ on at least $n-2$ players and to $O(N^\frac{n-2}{2})$ whenever the inputs in the correlation Bell inequality is at most $N$ for at least $n-2$ players.

The key point to prove Theorem \ref{Unbounded Bell violation} is the use of random states, showing once more that these states exhibit unexpected extremal properties. The proof of the theorem relies on hard techniques from operator spaces and it is highly nonconstructive. In particular, it does not provide neither an explicit (nor even probabilistic) form of the Bell inequality $T$ attaining such a violation nor any control on the dimension $D$ appearing in the statement of the theorem. Since we will explain below a more recent result improving Theorem \ref{Unbounded Bell violation}, we will not say too much about the proof of the previous result. Instead, we will present some basic ideas in order to be able to discuss the analogies and differences with the later proof.

The first idea in the proof of Theorem \ref{Unbounded Bell violation} is to reduce the problem to work with Hilbert spaces. Indeed, the initial problem consists of comparing two difficult to handle quantities $\omega(T)$ and $\omega^*(T)$ for a given tensor $T=(T_{i,j,k})_{i,j,k=1}^N$. Let us consider another tensor $S=(S_{i,j,k})_{i,j,k=1}^m$ as an element in $\mathbb C^m\otimes \mathbb C^m\otimes \mathbb C^m$ and define two new quantities (norms):
$$\|S\|_{\ell_2^m\otimes_\epsilon \ell_2^m\otimes_\epsilon\ell_2^m}=\sup\Big\{\Big|\sum_{i,j,k=1}^mS_{i,j,k}a_ib_jc_k\Big|: \|(a_i)_{i=1}^m\|_2, \|(b_j)_{j=1}^m\|_2, \|(c_k)_{j=1}^m\|_2\leq 1\Big\}$$and 
$$\|S\|_{*}=\sup\Big\{\Big\|\sum_{i,j,k=1}^mS_{i,j,k}A_i\otimes B_j\otimes C_k\Big\|_{M_{d^3}}: \|(A_i)_{i=1}^m\|_{RC}, \|(B_j)_{j=1}^m\|_{RC}, \|(C_k)_{k=1}^m\|_{RC}\leq 1\Big\},$$where for a sequence of complex numbers $(z_j)_{j=1}^m$, $\|(z_j)_{j=1}^m\|_2$ is the euclidean norm and for a sequence of matrices $(Z_j)_{j=1}^m\subset M_d$ we define
\begin{align}\label{RC norm}
\|(Z_j)_{j=1}^m\|_{RC}=\max\Big\{\big\|\sum_{j=1}^mZ_jZ_j^\dag\big\|^{\frac{1}{2}}, \big\|\sum_{j=1}^mZ_j^\dag Z_j\big\|^{\frac{1}{2}}\Big\}.
\end{align}These quantities can be understood as a \emph{hilbertian version} of the values $\omega(\cdot)$ and $\omega^*(\cdot)$. On the other hand, for every tensor $S$ one can construct another tensor $T=(T_{i,j,k})_{i,j,k=1}^N$ with $N=2^m$ for which, up to a universal (known) constant, $$\frac{\omega^*(T)}{\omega(T)}\simeq \frac{\|S\|_{*}}{\|S\|_{\ell_2^m\otimes_\epsilon \ell_2^m\otimes_\epsilon\ell_2^m}}.$$

Here, we should point out that one could explicitly construct both the Bell inequality $T$ and the observables to be used to compute $\omega^*(T)$ from the elements $S$, $(A_i)_{i=1}^m$, $(B_j)_{j=1}^m$, $(C_k)_{k=1}^m$ used to compute $\|S\|_{*}$. In order to find a tensor $S$ for which the quotient between $\|S\|_{*}$ and  $\|S\|_{\ell_2^m\otimes_\epsilon \ell_2^m\otimes_\epsilon\ell_2^m}$ is large, one considers the random state $|\psi\rangle=\frac{1}{m}\sum_{i,j,k=1}^mU_i(j,k) |ijk\rangle\in \mathbb (\mathbb C^{m})^{\otimes_3}$. It follows from well known results in random matrices that  $\langle\psi|\psi\rangle\simeq 1$. Then, by doubling indices one can naturally consider the element $S$ in $\mathbb C^{m^2}\otimes \mathbb C^{m^2}\otimes \mathbb C^{m^2}$ defined by $S_{i,i';j,j';k,k'}=\langle k|U_i^{tr}|j\rangle\langle k'|U_{i'}^\dag|j'\rangle$ and the matrices $A_{i,i'}=\frac{1}{\sqrt{m}}|i\rangle\langle i'|$, $B_{j,j'}=\frac{1}{\sqrt{m}}|j\rangle\langle j'|$, $C_{k,k'}=\frac{1}{\sqrt{m}}|k\rangle\langle k'|$ for every $i,i',j,j',k,k'=1,\cdots, m$. It is not difficult to check that for  these matrices the quantity (\ref{RC norm}) is lower than or equal to one, which implies
\begin{align}\label{* norm lower bound}
\|S\|_{*}&\geq \frac{1}{m^{\frac{3}{2}}}\Big|\Big\langle \psi\Big| \sum_{i,i',j,j',k,k'=1}^mS_{i,i',j,j',k,k'}|i\rangle\langle i'|\otimes |j\rangle\langle j'|\otimes |k\rangle\langle k'| \Big|\psi\Big\rangle\Big|\\\nonumber&
=\frac{1}{m^{\frac{7}{2}}}\sum_{i,i'=1}^mtr\Big(\big(\bar{U_i}\otimes U_{i'}^\dag\big)\big(U_i^{tr}\otimes U_{i'}\big)\Big)=\frac{1}{m^{\frac{3}{2}}}tr\big(\uno_{M_{m^2}}\big)=\sqrt{m}.
\end{align}

According to the previous comments, we would finish the proof if we could show that $\|S\|_{\ell_2^{m^2}\otimes_\epsilon \ell_2^{m^2}\otimes_\epsilon \ell_2^{m^2}}\lesssim 1$. Unfortunately, it turns out that this estimate is not true. The key result \cite[Proposition 20]{PWPVJ} shows that for every $\epsilon>0$ and for every $m\in \mathbb N$, there exist $N(\epsilon, m)\in \mathbb N$, some unitary matrices $U_i\in M_N$ and some matrices $H_{i,i'}\in M_{N^2}$ for every $i,i'=1,\cdots , m$ such that if we define $F_{i,i'}=U_i^{tr}\otimes U_{i'}^\dag+H_{i,i'}$ and $S_{i,i';j,j';k,k'}=\langle kk'|F_{i,i'}|jj'\rangle$, we have that
\begin{align}\label{key conditions}
\|S\|_{\ell_2^{m^2}\otimes_\epsilon \ell_2^{N^2}\otimes_\epsilon \ell_2^{N^2}}\lesssim 1\text{       } \text{    and     }\text{       }\big|\frac{1}{N^2}tr\big(H_{i,i'}(\bar{U_i}\otimes U_{i'})\big)\big|<\epsilon \text{       } \text{    for every     } i,i'=1,\cdots ,m.
\end{align}This means that, while the value of the norm $\|\cdot\|_{\ell_2^{m^2}\otimes_\epsilon \ell_2^{N^2}\otimes_\epsilon \ell_2^{N^2}}$ is smaller for the new inequality $S$, the modification with respect to the previous inequality does not affect essentially to the estimate (\ref{* norm lower bound}). Hence, the result follows. As the reader can guess, it is precisely in the proof of the existence of these highly nontrivial matrices $H_{i,i'}$ where the explicitness of the result is lost. 

In \cite{Pisier I}, \cite{Pisier II} the author refined the preceding proof by using random gaussian matrices instead of random unitaries. The problem is again reduced to separate the norms $\|S\|_{\ell_2^m\otimes_\epsilon \ell_2^m\otimes_\epsilon\ell_2^m}$ and $\|S\|_{*}$ but the technical proof to find the previous unitaries is replaced by the use of a previous known result (\cite[Theorem 16.6]{Pisier I}), which seems to be due to Steen Thorbjørnsen. This result guarantees the existence of a family of $N\times N$ gaussian matrices $(G_i)_{i=1}^m$ for a certain $N$, for which the construction explained above can be done directly. The main advantage of this approach is that one can use directly these gaussian matrices and the matrices $H_{i,i'}$ are not needed anymore. Actually, one can follow the estimates behind these results and replace the previous transformation from the tensor $S$ to $T$ by a more sophisticated one (allowing $N\approx n^2$) to obtain a Bell inequality $T=(T_{i,j,k})_{i,j,k=1}^{N^4,N^8,N^8}$ which can give violations of order $\sqrt{N}$ by using a quantum state in $\mathbb C^N\otimes \mathbb C^{N^2}\otimes \mathbb C^{N^2}$ (see \cite[Remark 3.3]{Pisier II} for details). 
\subsection{Briet and Vidick's construction}
In the work \cite{BrVi} the authors gave another proof of Theorem \ref{Unbounded Bell violation} which considerably improved both the estimates on the parameters and the construction.
\begin{theorem}\label{theorem BV}
Let us assume that $N=2^j$ for some natural number $j$. There exist a quantum pure states $|\psi\rangle$ in $\mathbb C^N\otimes \mathbb C^N\otimes \mathbb C^N$ and a Bell inequality $T=(T_{i,j,k})_{i,j,k=1}^{N^2,N^2,N^2}$ such that the violation by $|\psi\rangle$ on such an inequality is $\Omega\big(  \sqrt{N}\log^{-5/2} N\big)$. Moreover, the observables used in each party are tensor products of Pauli matrices.

More generally, for every $N$ that is a power of 2 there exist an $n$-party Bell inequality $T$ with $N^2$ inputs in each party and a state $|\psi\rangle\in \mathbb  (\mathbb C^N)^{\otimes n}$ such that they can lead to a violation $\Omega\big( (N\log^{-5} N)^{\frac{n-2}{2}}\big)$, where the observables used in each party are tensor products of Pauli matrices. 
\end{theorem}
According to Theorem \ref{upper bounds tripartite} and the comments below it, the previous result is optimal, up to a logarithmic factor\footnote{Pisier has shown in \cite{Pisier II} that such a factor can be reduced to $\log ^{-\frac{3}{2}} N$.}, in the dimension of the Hilbert spaces and it is only quadratically off from the best upper bound as a function of the number of inputs $N$. The proof of Theorem \ref{theorem BV} is probabilistic, being based again on the construction of a random Bell inequality which interacts properly with a random pure state via some suitable observables. In fact, since the elements involved in Theorem \ref{theorem BV}  are very simple, we will first explain them and a brief explanation at a more mathematical level will be discussed later.

Let us consider the random state
\begin{align}\label{gaussina state}
|\varphi\rangle=\frac{1}{\|\bar{g}\|}\sum_{i,j,k=1}^Ng_{i,j,k}|i,j,k\rangle,
\end{align}where $\bar{g}$ is the corresponding non-normalized state. A key point in \cite{BrVi} is the use of Pauli matrices\footnote{The only property used by the authors is that the tensor products of $j$ Pauli matrices form an orthogonal  basis of $M_N$ formed by observables. Any other such a system would equally work in the proof.} to show that if one considers a six indices tensor $(S_{i,i';j,j';k,k'})_{i,i',j,j',k,k'=1}^N$, one can easily define a Bell inequality $T=(T_{P,Q,R})_{P,Q,R}$, where the inputs are indexed in the tensor product of $j$ Pauli matrices and such that if the three parties use respectively the observables $P$, $Q$ and $R$ associated to the inputs $P$, $Q$ and $R$, then it turns out that
\begin{align}\label{3-3 norm}
\sum_{P,Q,R}T_{P,Q,R}\langle \varphi |P\otimes Q\otimes R|\varphi\rangle= N^3\langle \varphi|S|\varphi\rangle.
\end{align}Here $S=\sum_{i,i';j,j';k,k'=1}^NS_{i,i';j,j';k,k'}|ijk \rangle \langle i'j'k' |$  is regarded as an element in $M_{N^3}$. Indeed, this follows easily from the fact that the set $\mathcal P_j=\{\text{Pauli matrices}\}^{\otimes_j}$ forms an orthogonal basis of $M_N$ with respect to the inner product defined as $\langle A, B\rangle=tr(AB^\dag)$. Moreover, it is trivial to check that $\langle P, Q\rangle=N\delta_{P,Q}$ for every $P,Q\in \mathcal P_j$. Then, by considering the set $\mathcal P_j\otimes \mathcal P_j\otimes \mathcal P_j$ one obtains an orthogonal basis of $M_{N^3}$, and for every element $S$ in this space one can write (its Fourier expansion) $$S=\frac{1}{N^3}\sum_{P,Q,R\in \mathcal P_j}\langle S, P\otimes Q\otimes R\rangle P\otimes Q\otimes R.$$Hence, if one defines $T=(T_{P,Q,R})_{P,Q,R\in \mathcal P_j}$ such that $$T_{P,Q,R}=\langle S, P\otimes Q\otimes R\rangle=\sum_{i,i';j,j';k,k'=1}^NS_{i,i';j,j';k,k'}P_{i,i'}Q_{j,j'}R_{k,k'},$$ one has the desired property (\ref{3-3 norm}). Note that in order to obtain this property one needs to double indices as it was made in the proof of Theorem \ref{Unbounded Bell violation}.

On the other hand, let us assume that $\chi, \nu, \zeta:\mathcal P_j\rightarrow \{-1,1\}$ are optimal functions to compute $\omega(T)$. By defining the hermitian matrices $X=\sum_{P\in \mathcal P_j}\chi(P)P$, $Y=\sum_{Q\in \mathcal P_j}\nu(Q)Q$, $Z=\sum_{R\in \mathcal P_j}\zeta(R)R$ in $M_N$ one can write
\begin{align*}
\omega(T)=\sum_{P,Q,R\in \mathcal P_j}T_{P,Q,R}\chi(P)\nu(Q)\zeta(R)=\langle S, X\otimes Y\otimes Z\rangle.
\end{align*}Then, the fact $\|X\|_2=\|Y\|_2=\|Z\|_2=N^{3/2}$ implies that
\begin{align}\label{2-2-2 norm}
\omega(T)\leq N^{9/2}\sup_{X,Y,Z\in B(\text{Herm}(N))}\langle S, X\otimes Y\otimes Z\rangle,
\end{align}where $B(\text{Herm}(N))$ denotes the unit ball of the hermitian $N\times N$ matrices with respect to the Frobenius norm.

By looking at Eq. (\ref{3-3 norm}), a potential good choice for the  tensor $S$ is $S_{i,i';j,j';k,k'}=g_{i,j,k}g_{i',j',k'}$ for every $i,i',j,j',k,k'$. Indeed, if one considers this element it is straightforward to check that
\begin{align}\label{Hilbertian norm Gaussian}
\omega^*(T)\geq N^3\langle \varphi|S|\varphi\rangle=N^3\|\bar{g}\|_2^2\simeq N^6 \text{    }  \text{    }  \text{  with high probability over $g$},
\end{align}where the last estimate follows from well known results on gaussian variables. Therefore, the statement would follow if one proved that the quantity $\sup_{X,Y,Z\in B(\text{Herm}(N))}\langle T, X\otimes Y\otimes Z\rangle$ is upper bounded by $N$ up to, maybe, some logarithmic terms. Unfortunately, it can be deduced from known results that the previous amount is $\Omega(N\sqrt{N})$. The key point in the construction by Briet and Vidick is to remove some of the indices of $S$. More precisely, they consider the tensor $S$ defined by $S_{i,i';j,j';k,k'}=g_{i,j,k}g_{i',j',k'}$ whenever $i\neq i'$, $j\neq j'$, $k\neq k'$ and $S_{i,i';j,j';k,k'}=0$ otherwise. It is very easy to see that the corresponding values $\omega^*(T)$ has the same order $N^6$, since the number of removed terms is negligible in the estimate (\ref{Hilbertian norm Gaussian}). On the other hand, the main result in \cite{BrVi} shows that the classical value of the new inequality does decrease by the previous modification:
\begin{align}\label{key estimate BV}
\sup_{X,Y,Z\in B(\text{Herm}(N))}\langle S, X\otimes Y\otimes Z\rangle\lesssim N \log ^{5/2} N \text{    }  \text{    }  \text{  with high probability over $g$.}   
\end{align}This estimate allows to obtain Theorem \ref{theorem BV}. Note that here, as in the proof of Theorem \ref{Unbounded Bell violation}, the definition of the Bell inequality $T$ can lead to complex coefficients $T_{P,Q,R}=\sum_{i,i';j,j';k,k'=1}^NT_{i,i';j,j';k,k'}P_{i,i'}Q_{j,j'}R_{k,k'}$. However, it is trivial that either the real part or the imaginary part must lead to a similar violation up to a constant $2$.

The proof of Eq. (\ref{key estimate BV}) is based on a technical $\epsilon$-net construction using a decomposition of elements in $B(\text{Herm}(N))$ as linear combinations of normalized projections. In their proof the authors show that  the corresponding estimate holds for such projections with a sufficiently good concentration so that, by applying a counting argument on the elements of the net, they obtain the result.

A more careful look at the previous argument allows to see some analogies with the proof of Theorem \ref{Unbounded Bell violation} (and the subsequent improvement in \cite{Pisier II}). Indeed, one can see that the proof by Briet and Vidick is also reduced to separate two norms in $\mathbb C^{m^2}\otimes \mathbb C^{m^2}\otimes \mathbb C^{m^2}$, being the first one again the $\epsilon$ norm $\|S\|_{\ell_2^{m^2}\otimes_\epsilon \ell_2^{m^2}\otimes_\epsilon\ell_2^{m^2}}$. However, in this case the second norm is:
$$\|S\|_{3,3}=\big\|\sum_{i,i';j,j';k,k'}^mS_{i,i';j,j';k,k'}|i,j,k\rangle\langle i',j',k'\rangle\big\|_{M_{N^3}}.$$The reader should note that, while in the proof of Theorem \ref{Unbounded Bell violation} the comparison between the norms consisted of computing the norm of the identity map on $\otimes_{i=1}^3\mathbb C^{m^2}$ when this space is endowed with the $\epsilon$ and the $\|\cdot\|_{*}$ norm respectively, in the proof of Theorem \ref{theorem BV} one must compute the norm of the rearrangement map $R: \otimes_{i=1}^3\mathbb C^{m^2}\rightarrow \otimes_{i=1}^2\mathbb C^{m^3}$ defined by $R(|ii'\rangle|jj'\rangle|kk'\rangle) = |ijk\rangle|i'j'k'\rangle$. Note that doubling indices is an essential point to define this map.

There are two key points to understand the improvement in Theorem \ref{theorem BV} with respect to the previous results. The first one is that the transformation from the tensor $S$ in $\mathbb C^{m^2}\otimes \mathbb C^{m^2}\otimes \mathbb C^{m^2}$ into the Bell inequality $T=T(S)$ does not imply an increase in the number of inputs. That is, one has $m=N$. The second  point is that the authors gave an essentially optimal separation from the norms they considered in $\mathbb C^{m^2}\otimes \mathbb C^{m^2}\otimes \mathbb C^{m^2}$ (see \cite[Section 2]{Pisier II}  for details). Finally, we should mention that Theorem \ref{theorem BV} is not just an existence result, but the proof shows that such an estimate happens with high probability in the choice of the gaussian variables $g$. 

Theorem \ref{theorem BV} can be understood as a result proving that tripartite correlation Bell inequalities give large violations with high probability when they are properly sampled. However, this sampling is rather artificial since, in particular, it implies doubling indices in the coefficients. Regarding Section \ref{Section: Bipartite correlation Bell inequalities} in this survey, a next step in the problem could be to study the behavior of $T=(\epsilon_{i,j,k})_{i,j,k=1}^N$, where it is sampled from $\{-1,1\}^{N^3}$ according to the uniform measure. In fact, this is equivalent to consider $T=(g_{i,j,k})_{i,j,k=1}^N$, a family of independent real gaussian variables. It is not difficult to see that $\mathbb E \big[\omega(T)\big]\lesssim N^2$ in this case. Any nontrivial estimate on $\mathbb E \big[\omega^*(T)\big]$ would definitely be an interesting result. On the other hand, despite the sharpness of Theorem \ref{theorem BV} there is still a gap with respect to the best known upper bound for the multipartite Bell violation as a function of the number of inputs. It would be interesting to know if one can indeed attain violation of order $\sqrt{N}$ by using a tripartite correlation Bell inequality with $N$ inputs per party. Finally, any explicit (non-probabilistic) construction of a tripartite correlation Bell inequality leading to unbounded violations would be also very welcome by the community.
\section{Correlation Bell inequalities for a large number of parties}\label{Section: Correlation Bell inequalities for a large number of parties}
\subsection{Symmetric XOR games}\label{Section: Symmetric XOR games}
In the works \cite{Ambainis II}, \cite{Ambainis III} and \cite{Ambainis IV}  Ambainis and coauthors studied the scenario of binary inputs correlation Bell inequalities with \emph{many} parties $T=(T_{x_1,\cdots ,x_n})_{x_1,\cdots ,x_{n}\in\{0,1\}}$. In this case, it is well known that the quotient between $\omega^*(T)$ and $\omega (T)$ can be equal to $2^\frac{n}{2}$ for some particular inequalities $T$ (\cite{Ardehali},  \cite{Mermin}). Moreover, it is also known that such a violation is optimal in this setting \cite{WeWo I}, \cite{WeWo II}.  The reader should immediatly note that this problem is completely different from the one considered in the previous section, where unbounded Bell violations were proved by considering only a fixed number of parties $n$ (say three) and increasing the number of inputs $N$. In order to clarify the connection between the works \cite{Ambainis II}, \cite{Ambainis III}, \cite{Ambainis IV} and the explanation in this survey, let us recall the reader that any correlation Bell inequality $T=(T_{x_1,\cdots, x_n})_{x_1,\cdots, x_n}$ verifying $\sum_{x_1,\cdots, x_n}|T_{x_1,\cdots, x_n}|=1$ can be trivially written as $T=\big(\pi(x_1,\cdots, x_n)f(x_1,\cdots, x_n)\big)_{x_1,\cdots, x_n}$, where $\pi$ is a probability distribution over the set of inputs and $f:X_1\times \cdots \times X_N\rightarrow \{-1,1\}$ is a function. This gives a correspondence between the classical (resp. quantum) value of a correlation Bell inequality and the classical (resp. quantum) bias (probability of winning minus probability of loosing) of an XOR game. The corresponding game is defined by the distribution $\pi$ over the set of questions (inputs) and the predicate function, given by $V(x_1,\cdots, x_n, a_1,\cdots , a_n)=1$ if and only if $a_1\oplus\cdots \oplus a_n=\frac{1}{2}(f(x_1,\cdots, x_n)+1)$. Then, a \emph{symmetric XOR game} is a correlation Bell inequality where $\pi$ is the uniform probability distribution over the set of inputs and the function $f$ is invariant under permutations of the parties\footnote{We will not talk about symmetric correlation Bell inequalities since these should be defined as those inequality which are invariant under permutations of the parties without any extra restriction.}. Note that for the case of binary inputs $x_i=0,1$, a symmetric XOR game $T$ can be described by means of a sequence of $n+1$ bits $(M_0,M_1,\cdots, M_n)$, where $M_j=f(x_1,\cdots, x_n)$ whenever $\sum_{i=1}^nx_i=j$. Then, one can define a probability distribution on the set of these Bell inequalities by picking the $(n+1)$-bit string $(M_0,M_1,\cdots, M_n)$ uniformly at random. The main results in \cite{Ambainis II} and \cite{Ambainis III} imply that for every $\epsilon>0$ there exist positive constants $C_1(\epsilon)$, $C_2(\epsilon)$ such that for a high enough $n$ we have
\begin{align}\label{violation symmetric}
\mathbb P\Big\{T:\frac{\omega^*(T)}{\omega (T)}\in \Big[C_1(\epsilon)\sqrt{\log n}, C_2(\epsilon)\sqrt{\log n}\Big]\Big\}\geq 1-\epsilon.
\end{align}
Therefore, large Bell violations occur with probability very close to one on the set of all symmetric XOR games. On the other hand, since the violation $2^\frac{n}{2}$ mentioned before is attained on symmetric XOR games, one sees that the violation $\sqrt{\log n}$ is far from being optimal.

The key point to prove Eq. (\ref{violation symmetric}) is a simplification of both quantities $\omega(T)$ and $\omega^*(T)$, when $T$ is a symmetric XOR games \cite{Ambainis IV}. On the one hand, in \cite{Ambainis III} the authors proved that in order to study $\omega(T)$ in this case, it suffices to look at $n+1$ deterministic correlations $\gamma_0, \cdots, \gamma_n\in \mathcal L$. These correlations are very easy to describe in terms of the strategy followed by the players to play the XOR game corresponding to $T$. Indeed, $\gamma_k$ is the correlation obtained if the players follow the deterministic strategy denoted by $(00)^k(01)^{n-k}$, $k=0,\cdots, n$. Here, for a fixed $k$ the previous notation means that the first $k$ players answer aways the output $0$ while the next $n-k$ players answer the output $0$ if they are asked question $0$ and they output $1$ if they are asked question $1$. Interestingly, one can show in addition that only two of these strategies ($k=0$ and $k=n$) are relevant in average. Indeed, \cite[Theorem 4]{Ambainis III} and \cite[Theorem 5]{Ambainis III} show, respectively, that
\begin{align}\label{classical symmetric I}
\mathbb E\Big\{\max\big\{\big|\langle T,\gamma_0\rangle\big|, \big|\langle T,\gamma_n\rangle\big|\big\}\Big\}=\frac{0.8475...+o(1)}{n^{\frac{1}{4}}},
\end{align}and
\begin{align}\label{classical symmetric II}
\mathbb P\Big\{\max_{1\leq k\leq n-1}\big|\langle T,\gamma_k\rangle\big|\geq \frac{c}{n^{\frac{1}{4}}}\Big\}=O\Big(\frac{1}{n}\Big) \text{      }\text{    for any constant    }c>0.
\end{align}
A simplification in the analysis of the classical value of a symmetric XOR game allows the authors in \cite{Ambainis III} to prove Eq. (\ref{classical symmetric I}) and Eq. (\ref{classical symmetric II}) by using classical probabilistic techniques. One can deduce from here that for every $\epsilon>0$ there exist positive constants $C_1(\epsilon)$, $C_2(\epsilon)$ such that the probability of the classical value $\omega(T)$ being in $[C_1(\epsilon) n^{-1/4}, C_2(\epsilon) n^{-1/4}]$ is at least $1-\epsilon$. This is done by using that the random variable  $\max\big\{\big|\langle T,\gamma_0\rangle\big|, \big|\langle T,\gamma_n\rangle\big|\big\}$ converges (as $n$ tends to infinity) to the sum of two Gaussian random variables with known mean and variance (see \cite[Section 5.2]{Ambainis II} for details). However, note that this does not allow to state the previous estimate for fixed constants $C_1$ and $C_2$ and probability $1-o(1)$. This lack in the concentration of measure seems to be due to the nature of the value $\omega(T)$ rather than to the analysis performed by the authors.

On the other hand, it was proved in  \cite[Theorem 1, Theorem 2]{Ambainis II} that
\begin{align}\label{quantum symmetric}
\lim_{n\rightarrow \infty}\mathbb P\Big[C_1\frac{\sqrt{\ln n}}{n^{\frac{1}{4}}}\leq \omega^*(T)\leq C_2\frac{\sqrt{\ln n}}{n^{\frac{1}{4}}}\Big]=1 \text{      }\text{    for certain constants    }C_1, C_2>0.
\end{align}We see that this result is, in terms of concentration, stronger than the previous one for the classical value of the game, since here we can fix the constants $C_1$, $C_2$ and get probability $1-o(1)$. To show Eq. (\ref{quantum symmetric}) the authors first used a previous result in \cite{WeWo I}, \cite{WeWo II} stating that the quantum value of a symmetric XOR game is given by $\omega^*(M)=\max_{|z|=1}\Big|\frac{1}{2^n}\sum_{j=0}^n(-1)^{M_j}\binom{n}{j}  z^j\Big|$. Then, computing $\omega^*(M)$ reduces to the maximization of the absolute value of a polynomial in one complex variables. On the other hand, the previous expression is trivially upper and lower bounded by functions depending on the real and imaginary part of the corresponding polynomial,
\begin{align*}
\max_{\alpha\in [0,2\pi]}\Big|\frac{1}{2^n}\sum_{j=0}^n(-1)^{M_j}\binom{n}{j}  \cos (j\alpha)\Big|,\text{       }\text{       }\max_{\alpha\in [0,2\pi]}\Big|\frac{1}{2^n}\sum_{j=0}^n(-1)^{M_j}\binom{n}{j}  \sin (j\alpha)\Big|.
\end{align*}
If $(M_0,M_1,\cdots, M_n)$ are chosen at random, these expressions reduce to random trigonometric polynomials studied in \cite{SaZy} and the problem is actually reduced to study the quantities $M_n(t)=\max_{\alpha\in [0,2\pi]}\big|P_n(x,t)\big|$ for $P_n(x,t)=\sum_{m=0}^nr_m\psi_m(t)\cos mx$, where $(\psi_m)_m$ is the Rademacher system. Although this is the key object of study in \cite{SaZy}, the results in this work do not fit directly in the problem considered in \cite{Ambainis II} and obtaining the estimate (\ref{quantum symmetric}) requires a careful study of the particular case.
\subsection{Sampling quantum $n$-partite states}
A different way of studying Bell violations consists of sampling quantum states according to some random procedure and looking at the violations they can produce. Note that this is not the same as sampling quantum correlations at random as it was considered in Section \ref{Section: Bipartite correlation Bell inequalities}. In the current situation one should regard both the POVMs and the Bell inequalities as free parameters in the problem. One can then define a natural measure of how nonlocal a quantum state is (see \cite{Palazuelos II}). However, the great freedom in the problem makes it difficult to handle. An intimately related problem is to study the probability of a quantum state being nonlocal with respect to a particular family of Bell inequalities. This is the context considered in \cite{DrOl}, where the authors restricted to the setting of $n$-partite binary inputs correlation Bell inequalities $(T_{x_1,\cdots ,x_n})_{x_1,\cdots ,x_{n}\in\{0,1\}}$. It is well known that in this case the set of correlation Bell inequalities\footnote{Here, we really mean the equations defining the facets of the set $\mathcal L$.} can be replaced by the single nonlinear inequality \cite{WeWo II}, \cite{ZuBr}
\begin{align*}
\sum_{1\leq i\leq n}\sum_{x_i\in \{0,1\}}\Big|\prod_{j=1}^n\frac{A_0^j+(-1)^{x_j}A_1^j}{2}\Big|\leq 1,
\end{align*}where $A_0^j, A_1^j=\pm 1$, $1\leq j\leq n$ represent the deterministic measurement results for the pair of measurements in site $j$. Hence, one can define the function $LV:\mathbb S^{2d^n-1}\rightarrow \mathbb R^+$ on the unit sphere of $(\mathbb C^d)^{\otimes n}$; that is, the set of pure quantum state $|\psi\rangle$ of $n$ $d$-dimensional systems, by\begin{align*}
LV({|\psi\rangle}):=\sup \sum_{1\leq i\leq n}\sum_{x_i\in \{0,1\}}\Big\langle \psi\Big|\bigotimes_{j=1}^n\frac{A_0^j+(-1)^{x_j}A_1^j}{2}\Big|\psi\Big\rangle.
\end{align*}Here, the supremum runs over all possible choices of pair of observables $A_0^j$,  $A_1^j$ per site $j=1,\cdots ,n$. The main result in \cite{DrOl} describes the probabilistic behavior of $LV$ with respect to the uniform measure on $\mathbb S^{2d^n-1}$.
\begin{theorem}\label{Theorem Oliveira concentration}
Let $n$ and $d$ be two natural numbers larger than 1, let $|\psi\rangle$ be a unit vector distributed according to the uniform measure in the unit sphere $\mathbb S^{2d^n-1}$ of $(\mathbb C^d)^{\otimes n}$ and denote $\mathcal A_v=\{|\psi\rangle: LV_{|\psi\rangle}> v\}$. Then, the following inequality holds:
\begin{align}\label{Oliveira concentration}
\mathbb P (\mathcal A_v)\leq 2
\Big(\frac{n2^{n+1}d^2}{\delta}+2\Big)^{2d^2n}e^{-\frac{(v-\delta-c_{d,n})^2(\frac{d}{2})^n}{9\pi^3}}.
\end{align}Here, $\delta$ is any positive number, $v>c_{d,n}+\delta$ and $c_{d,n}=(\frac{2}{d})^\frac{n}{2}+\frac{d-2}{2}$.
\end{theorem}
As we mentioned in the previous section, for binary inputs correlation Bell inequalities the optimal quotient between the quantum and the classical value is $2^{\frac{n}{2}}$. Note that Eq. (\ref{Oliveira concentration}) implies that for large $n$ most pure states do not get even close to this violation. For the particular case $d=2$ (qubits) one has that $c_{d,n}=1$ for every $n$, so as long as $v\geq cn$ for a suitable constant $c$, $\mathbb P (\mathcal A_v)\rightarrow 0$ as $n$ goes to infinity. It was conjectured in \cite{Pitowsky} that this is also true if $v$ is of order $\sqrt{n\log n}$ but this problem seems to remain open (see \cite[Section I]{DrOl} for details). On the other hand, if $d\geq 3$ Eq. (\ref{Oliveira concentration}) implies that most of the states do not violate any of these inequalities for $n$ large enough. Indeed, note that $\lim_{n}c_{d,n}=\frac{d-2}{2}<1$, so we can choose $\delta>0$ with $\delta+c_{d,n}<v<1$ such that $\lim_{n}\mathbb P (\mathcal A_v)=0$ super exponentially. It could be interesting to study this problem when one samples on not necessarily pure states. Although the way of sampling in that case is not so obvious, there are very interesting works showing different behaviors in that picture (see \cite{ASY}, \cite{SWZ}, \cite{ZPNC} and the references therein).

Theorem \ref{Theorem Oliveira concentration} is proved via a concentration-type argument. First, the authors consider an $\epsilon$-net for the set $\mathcal Q$ of possible pairs of observables $A_0^j$,  $A_1^j$ per site $j=1,\cdots , n$. This allows them to fix a finite set $\mathcal Q_\epsilon$ of these elements to work with. On the other hand, for one such a choice $Q\in \mathcal Q_\epsilon$, one considers the corresponding function $LV^Q({|\psi\rangle}):\mathbb S^{2d^n-1}\rightarrow \mathbb R^+$. It turns out that this function is quite regular in terms of its Lipschitz constant. Then, one uses this regularity in two different ways. First of all, it allows to pass from taking the supremum on $\mathcal Q$ to considering $\mathcal Q_\epsilon$ by means of the inequality $$\mathbb P (\mathcal A_v):=\mathbb P \big\{|\psi\rangle: \sup_{Q\in \mathcal Q}LV^Q{|\psi\rangle}> v\big\}\leq \mathbb P \big\{|\psi\rangle:\sup_{Q\in \mathcal Q_\epsilon}LV^Q(|\psi\rangle)>v-\delta\big\}.$$Secondly, for a fixed $Q\in \mathcal Q_\epsilon$ one can compute the expectation $\mathbb E[LV^Q]$ and use Levy's Lemma to bound the probability of the points for which the function is far from its expectation $\mathbb P\big(|f-E[f]|>\epsilon\big)\leq 2\exp\big(-\frac{(n+1)\epsilon^2}{9\pi^3\lambda}\big)$, where here $f:\mathbb S^n\rightarrow \mathbb R$ is a real function and $\lambda$ is its Lipschitz constant with respect to the euclidean distance. Putting the two previous ingredients together allows the authors to obtain Eq. (\ref{Oliveira concentration}) by using a counting argument.
\subsection{Sampling random measurements}
In \cite{LHBR}, \cite{SVLBO}, \cite{WLB},  the authors considered quantum correlations sampled in such a way that the quantum state is fixed and the observables are chosen at random. This seems to be a very interesting problem from an experimental point of view since it is motivated by the requirement of well calibrated devices and a common reference frame between the parties in the standard Bell experiments. Showing that random measurements produce Bell violations implies that the previous assumptions can be removed. In \cite{LHBR}, \cite{WLB} the authors considered the $n$-partite $2$-dimensional GHZ state: $|\psi\rangle=2^{-1/2}\sum_{i=1}^2|i\rangle^{\otimes n}$ and studied the correlations when each party $j$ measures with two possible observables $A_{1,j}$, $A_{2,j}$ chosen at random according to the law (RIM): $A_{i,j}=n_{i,j}\cdot \sigma$, where $n_{i,j}$ are uniform and independent vectors on the unit sphere of $\mathbb C^3$ and $\sigma=(\sigma_x,\sigma_y, \sigma_z)$ is defined via the Pauli matrices. Numerical computations suggest that the corresponding correlations are nonlocal with probability tending to one as $n$ goes to infinity. This behavior is also studied when the two measurement per party are chosen so that $n_1^j$ and $n_2^j$ are orthogonal but still random (ROM) to show that nonlocality is even more likely in this last case. In \cite{SVLBO} the authors also showed some numerical evidences that in the bipartite case, correlations produced by using $N$ random measurements on each party of $|\psi\rangle=2^{-1/2}\sum_{i=1}^2|i\rangle^{\otimes 2}$ lead to a similar behavior to the previous one. Since the techniques in these papers are a bit different from those considered in the current survey we will not explain them in detail. However, we refer the reader to the original works, where many other results can be found as well as some very nice explanations about the experimental interpretation of the different ways of sampling.
\section{Unbounded violations for bipartite Bell inequalities}\label{General bipartite Bell inequalities}
The setting of general Bell inequalities is much richer than the context of correlation Bell inequalities. The new parameter given by the number of outputs $K$ leads to study the quantity $LV_n(N,K,d)$ analogous to $LV_n(N,d)$ introduced in Section \ref{section Unbounded violation of tripartite Bell inequalities}, where $n$ is the number of parties. The application of the parallel repetition theorem \cite{Raz} to the magic square game (or any pseudo-telepathy game \cite{Brassard-review}) and also the work \cite{KRT} show that $LV_2(N,K,d)$ cannot be upper bounded by a uniform constant independent of the parameters $N$, $K$ and $d$ (see \cite[Introduction]{JP} for details). That is, there exist \emph{unbounded violations of bipartite Bell inequalities}. However, the previous unbounded Bell violations are far from the upper bounds provided in \cite[Theorem 14]{JP}.
\begin{theorem}\label{general upper bounds}
The following upper bound holds:
$$LV_2(N,K,d)=O\big(h\big), \text{      }    \text{  where  } \text{      }   h=\min \big\{N,K,d\big\}.$$
\end{theorem}

In the paper \cite{JPPVWI} the authors used a random construction to show the lower bound $LV_2([2^{\log^2N/2}]^N,N,N)=\Omega(\sqrt{N}/\log^2N)$, improving the previous estimates so far. Since this result has been recently improved in two different ways we will omit the details in this survey. In the current section we will explain a more recent result which improves and simplifies the work in \cite{JPPVWI}. Based on a random construction, in \cite{JP} the authors proved that $LV_2(N,N,N)=\Omega(\sqrt{N}/\sqrt{\log N})$. According to Theorem \ref{general upper bounds} the previous estimate is only quadratically off from the best upper bounds in all the parameters of the problem at the same time. Actually, the most relevant point of this result with respect to the previous ones is the decrease in the number of inputs from exponential to polynomial. We will also explain in Section \ref{Section: A comment about some sharp explicit constructions} the existence of a bipartite Bell inequality providing the lower bound $LV_2(2^{N}/N,N,N)=\Omega(N/\log^2 N)$. This gives an essentially optimal upper bound as a function of the number of outputs and the dimension of the Hilbert space. 
\subsection{Unbounded violation with polynomially many parameters}
In  \cite{JP} the authors introduced the following construction: Consider a fixed number of $\pm 1$ signs $\eps_{x,a}^k$ with $x,a,k=1,\cdots ,N$. For a constant $K$ one defines the vectors $|\tilde{u}_x^a\rangle=\frac{1}{\sqrt{NK}}(1, \epsilon_{x,a}^1,\cdots , \epsilon_{x,a}^N)$ in $\C^{N+1}$ for every $x,a=1,\cdots ,N$; and
\begin{enumerate}[leftmargin=*]
\item[a)] \emph{Bell inequality coefficients}:  $\big(\tilde{M}_{x,y}^{a,b}\big)_{x,y;a,b=1}^{N,N+1}$ given by $$\tilde{M}_{x,y}^{a,b}= \frac{1}{N^2}\sum_{k=1}^N\epsilon_{x,a}^k\epsilon_{y,b}^k$$ for $x,y,a,b=1,\cdots, N$ and $\tilde{M}_{x,y}^{a,b}=0$ otherwise.
\item[b)] \emph{POVMs measurements}: $\{\tilde{E}_x^a\}_{x,a=1}^{N,N+1}$ in $M_{N+1}$ as
   \[
    \tilde{E}_x^a \, = \,
    \begin{cases}   |\tilde{u}_x^a\rangle\langle \tilde{u}_x^a|
  & \mbox{ for }  a=1,\cdots ,N\, , \cr
  \mathbb{1}-\sum_{a=1}^n\tilde{E}_x^a & \mbox{ for }  a=N+1\,
  \end{cases} \]
  for $x=1,\cdots ,N.$
\item[c)] \emph{States}: Let $ |\varphi_\alpha\rangle= \sum_{i=1}^{N+1} \al_i |ii\rangle$, where $(\al_i)_{i=1}^{N+1}$ be a decreasing and positive sequence verifying $\sum_{i=1}^{N+1}\alpha_i^2=1$.
\end{enumerate}
The following result can be found in \cite[Theorem 2]{JP}.
\begin{theorem}\label{Min theorem JP}
There exist universal constants $C$ and $K$ such that for every natural number $N$ there exists a choice of signs $\{\epsilon_{x,a}^k\}_{x,a,k=1}^{N}$ verifying that $\{\tilde{E}_x^a\}_{x,a=1}^{N,N+1}$ define POVMs measurements,
$\omega(\tilde{M})\leq C\log N$ and
\begin{equation}\label{violation-constructive2}
\sum_{x,y; a,b=1}^{N,N+1}\tilde{M}_{x,y}^{a,b}\langle
 \varphi_\alpha|\tilde{E}_x^a\otimes \tilde{E}_y^b |\varphi_\alpha\rangle\geq
 \frac{1}{K^2} \,  \al_1\sum_{i=2}^{N+1} \al_i.
\end{equation}
Moreover, the probability of the elements (choices of signs) verifying Eq. (\ref{violation-constructive2}) when they are chosen independently and uniformly at random tends to $1$ exponentially fast as $N$ tends to infinity.
\end{theorem}
According to the previous theorem, by considering an asymmetric element of the form $\alpha_1=1/\sqrt{2}$ and $\alpha_i= 1/\sqrt{2N}$ one obtains  the estimate $LV(\tilde{M})=\Omega(\sqrt{N}/\log N)$ \footnote{It was shown in \cite{Regev} that $\omega(\tilde{M})=O(\sqrt{\log N})$ in Theorem \ref{Min theorem JP}, leading to $LV(\tilde{M})=\Omega(\sqrt{N/\log N})$.}. As in the case of tripartite bell inequalities, a probabilistic construction provides a good lower bound for $LV_2(N,N,N)$ which almost matches the known upper bounds. 

The estimate $\omega(\tilde{M})=O(\log N)$ in Theorem \ref{Min theorem JP} can be obtained from well known probabilistic results. The original proof in \cite{JP} was based on Chevet's inequality but it was improved and simplified in  \cite{Regev} leading to $\omega(\tilde{M})=O(\sqrt{\log N})$. This last proof is based on standard estimates on gaussian variables.

The key point to understand Eq. (\ref{violation-constructive2}) is to consider the families of rank one operators  $E_x^a=\frac{1}{\sqrt{N}}\big(|u_x^a\rangle\langle 0|\big)_{a=1}^N$ for every $x$, where $u_x^a=\frac{1}{\sqrt{NK}}( \epsilon_{x,a}^1,\cdots , \epsilon_{x,a}^N)$. Then, it is not difficult to see that if one defines the rank one operator $\eta=|00\rangle\langle \psi|$, where $|\psi\rangle=\frac{1}{\sqrt{N}}\sum_{i=1}^N|ii\rangle$, one obtains
\begin{align}\label{violation assymmetric}
\sum_{x,y;a,b=1}^N\tilde{M}_{x,y}^{a,b}tr(E_x^a\otimes E_b^y\eta)=\Omega(\sqrt{N}).
\end{align}The problem here is that the elements $(E_x^a)_{x,a=1}^{N}$ do not define a family of POVMs (the operators are not even positive!) and that the operator  $\eta$ is not a state. Nevertheless, the previous family of operators should be understood as a family of non-positive POVMs. The interesting point here is that the good properties of these operators assure that by replacing the vectors $u_x^a$ by $\tilde{u}_x^a$ as in the statement of Theorem \ref{Min theorem JP}, one obtains a new family of operators $\tilde{E}_x^a=|\tilde{u}_x^a\rangle\langle \tilde{u}_x^a|$ verifying $\sum_{a=1}^N\tilde{E}_x^a\leq \mathbb{1}$. The new elements $\tilde{E}_x^a$ are defined by adding more entries to the matrices defining the operators $E_x^a$ so that they are positive and verify the required property on their sum. In particular, $E_x^a$ is encoded in the first column of $\tilde{E}_x^a$ up to an extra element equal to $1$. Then, the reader can guess and easily check that if one considers the state $|\varphi \rangle=1/\sqrt{2} \big(|00\rangle+|\psi\rangle\big)$, the terms $|00\rangle\langle\psi|$ and $|\psi\rangle\langle 00|$ appearing in $\rho=|\varphi \rangle \langle\varphi |$ have the same effect as in Eq. (\ref{violation assymmetric}). Indeed, the rest of the terms in $\rho$ do not play any role. Actually, a similar argument can be used for a general state $ |\varphi_\alpha\rangle= \sum_{i=1}^{N+1} \al_i |ii\rangle$ to obtain the estimate (\ref{violation-constructive2}) (see \cite[Section 3]{JP} for details).

Finally, the reason to pass from $M$ to $\tilde{M}$ is that the the previous argument gives an element $\big(tr(\tilde{E}_x^a\otimes \tilde{E}_b^y\rho)\big)_{x,y,a,b=1}^{N}$ for which $\sum_{a=1}^N\tilde{E}_x^a\leq \mathbb{1}$ for every $x$. Hence, by naively adding an extra output $a=N+1$ per measurement $x$, the previous modification on the element $M$ allows one to complete the family of operators to obtain $\sum_{a=1}^{N+1}\tilde{E}_x^a= 1$. 

Interestingly, if one plugs the $(N+1)$-dimensional maximally entangled state $|\psi\rangle$ in Eq. (\ref{violation-constructive2}), no large violation is obtained. In fact, it is an open question whether one can get large Bell violations on $\tilde{M}$ by using the maximally entangled state, even if we do not restrict its dimension. In \cite{JP} the authors made a modification to the inequality $\tilde{M}$ to obtain such an example. More precisely, \cite[Theorem 3]{JP} shows the existence of a Bell inequality $\bar{M}$ with $2^{N^2}$ inputs and $N+1$ outputs par party, and POVMs $\{\bar{E}_x^a\}_{x,a}$ acting on $\mathbb C^{N+1}$ such that $\bar{M}$ and $\{\bar{E}_x^a\}_{x,a}$ verify the same properties as those in the statement of Theorem \ref{Min theorem JP} and, in addition, $\sup \{|\langle \bar{M},Q_{max}\rangle|\}=O(\log N)$, where this $\sup$ runs over all quantum probability distributions $Q_{max}$ constructed with the maximally entangled state in any dimension. Previous results in this direction had been obtained in \cite{LVB}, \cite{ViWe}, where the authors had proved that certain quantum probability distributions cannot be written by using the maximally entangled state. We will not explain \cite[Theorem 3]{JP} in detail since its proof is quite technical and it is based on proving certain estimates on completely bounded norms. Furthermore, an improvement of the previous construction was made in \cite[Section 3]{Regev}, where the author gave an explicit Bell inequality with $2^N/N$ inputs and $N$ outputs per party verifying the same properties as $\bar{M}$. The proof of this result is based on deep techniques from quantum information theory. 

As in the case of Theorem \ref{theorem BV}, one can understand the statement of Theorem \ref{Min theorem JP} as a result providing large Bell violations with high probability when the Bell inequalities are properly sampled. However, while in the case of tripartite correlations the sampling procedure was quite artificial, involving in particular a duplication of indices, in the current situation the sampling looks quite natural if one ignores the added extra zeros, which do not play an important role. It is not clear for the author if some other samplings should be considered more (or less) reasonable.

The lack in the sampling criterium becomes more extreme when one samples quantum probability distributions. This has motivated the works in this direction to focus on very particular situations. An example of this can be found in \cite{AtZo}, where the authors study the violation of the CGLMP bipartite Bell inequality $M_{CGMLP}^K$ with binary inputs and $K$ possible outputs per measurement \cite{CGLMP}. Actually, even though the inequality $M_{CGMLP}^K$ is fixed, the authors in \cite{AtZo} need an assumption about the best POVMs for that inequality to be able to analyze its probabilistic behavior. Once the measurements are fixed, the problem of sampling quantum probability distributions is reduced to sampling quantum states. The main result in \cite{AtZo} gives the expected value of the violation for the Bell inequality $M_{CGMLP}^K$ with respect to the uniform measure on $K$-dimensional bipartite pure states; that is, when the pure states are uniformly sampled from the unit sphere of $\mathbb C^{K^2}$. This is done analytically for $K=2$ and in the range of large $K$, while for intermediate values it is done numerically. The key point of this study is that the previous restrictions allow the author to write the quantum value of the corresponding inequality as a function of the Schmidt coefficients of the states (see \cite[Eq. (3)]{AtZo}). Interestingly, even for this restricted setting the authors need to use nontrivial techniques from random matrix theory to analyze the problem.
\subsection{A comment about some sharp explicit constructions}\label{Section: A comment about some sharp explicit constructions}
We will finish this work by briefly mentioning two fully explicit constructions introduced in \cite{BRSW}. Since this survey is focused on random constructions, we will not go on these examples in detail and we refer the reader to the original source. However, the relevance of these examples forces us to mention them. The first one, called \emph{Hidden matching game}, gives a bipartite Bell inequality with $2^N$ inputs and $N$ outputs per party, which leads to a violation of order $\sqrt{N}/\log N$. Although this order is slightly worse than the one in Theorem \ref{Min theorem JP} and some of the parameters are exponentially higher, this Bell inequality presents a very particular form. Indeed, it is introduced via a two-prover one-round game $G$ for which $\omega^*(G)=1$. However, we must point out that the authors in \cite{BRSW} show that the quotient between the quantum and the classical bias of the game $G$ is $\beta^*(G)/\beta(G)=\Omega(\sqrt{N}/\log N)$. From here one can obtain a Bell inequality $M$ (which is not a two-prover one-round game anymore and it has different quantum and classical values from $G$) such that $\omega^*(M)/\omega(M)=\Omega(\sqrt{N}/\log N)$ (see \cite[Section 2]{BRSW} for a discussion about these things). The second example provided in \cite{BRSW} is the \emph{KV game}, which was first introduced by Khot and Visnoi in \cite{KhVi} to show a large integrality gap for a SDP relaxation of certain complexity problems. In \cite{BRSW} the authors carried out a careful analysis of the game to show that it provides a Bell inequality with $2^N/N$ inputs and $N$ outputs per party, leading to a Bell violation of order $N/\log^2 N$. Since this order is attained on a quantum probability distribution constructed with the $N$-dimensional maximally entangled state, it follows that $LV_2(2^N/N,N,N)=\Omega(N/\log^2 N)$. According to Theorem \ref{general upper bounds}, the previous result is essentially optimal in both the number of outputs and the dimension of the Hilbert space. Finally, this Bell inequality is particularly interesting because it is a two-prover one-round game, which implies some additional structures (see \cite{BRSW} for details).  As an interesting remark we should point out that the fact that the maximally entangled state is the key element to study the KV game is not a coincidence. Indeed, it was proved in \cite[Theorem 10]{JP} that for any Bell inequality with positive coefficients (in particular, any two-prover one-round game) the maximally entangled state always gives the largest violation up to a logarithmic factor in the dimension of the Hilbert space. In this sense, the KV game is very different from the Bell inequality given in Theorem \ref{Min theorem JP}.

It would be interesting to study if one can obtain a similar violation to the one given by the KV game by reducing the number of inputs (at least from exponential to polynomial). Moreover, there is still a gap of order $\sqrt{N}$ between the best lower and upper bounds for the bipartite Bell violations as function of the number of inputs $N$. Finally, the quantity $LV_3(N,K,d)$ seems to be completely unexplored. One can check that some examples such as the one introduced in Theorem \ref{Min theorem JP} can be generalized to the multipartite setting to provide nontrivial lower bounds. Finding an analogous inequality to the KV game in the tripartite case seems to be a more challenging problem.
\section*{Acknowledgments}
Author's research was supported by the Spanish projects MTM2011-26912 and MINECO: ICMAT Severo Ochoa project SEV-2011-0087 and the ``Ram\'on y Cajal'' program. 


\begin{thebibliography}{99}
%
\bibitem{Ambainis I} A. Ambainis, A. Backurs, K. Balodis, D. Kravcenko, R. Ozols, J. Smotrovs, M. Virza, \emph{Quantum strategies are better than classical in almost any XOR game}, Automata, Languages, and Programming Lecture Notes in Computer Science Volume 7391, 25-37 (2012).  

\bibitem{Ambainis II} A. Ambainis, J. Iraids,  \emph{Provable Advantage for Quantum Strategies in Random Symmetric XOR Games}. Available in arXiv:1302.2347.

\bibitem{Ambainis III} A. Ambainis, J. Iraids, D. Kravchenko, M. Virza,  \emph{Advantage of quantum strategies in random symmetric XOR games}, Mathematical and Engineering Methods in Computer Science, Lecture Notes in Computer Science, vol. 7721, 57-68 (2013).

\bibitem{Ambainis IV} A. Ambainis, D. Kravchenko, N. Nahimovs, A. Rivosh, \emph{Nonlocal quantum XOR games for large number of players}, Theory and Applications of Models of Computation, Lecture Notes in Computer Science, vol. 6108, 72-83 (2010).

\bibitem{Ardehali} M. Ardehali, \emph{Bell inequalities with a magnitude of violation that grows exponentially with the number of particles}, Phys Rev A, 46(9), 5375-5378 (1992). 

\bibitem{AtZo} M.R. Atkin, S. Zohren, \emph{Violations of Bell inequalities from random pure states}. Available in arXiv:1407.8233.

\bibitem{ASY} G. Aubrun, S. Szarek, D. Ye,  \emph{Entanglement thresholds for random induced states}, Comm. Pure Appl. Math. 67, 129-171 (2014).

\bibitem{Bell}  J.S. Bell, \emph{On the Einstein-Poldolsky-Rosen paradox}, Physics 1, 195 (1964).

\bibitem{Blei}  R. C. Blei,  \emph{Multidimensional extensions of the Grothendieck inequality and applications}, Arkiv fur Matematik, 17, 51-68 (1979).

\bibitem{BPV}  F. Bombal, D. P\'{e}rez-Garc\'{\i}a, I. Villanueva, \emph{Multilinear extensions of Grothendieck's theorem}, Q. J. Math. 55, 441-450 (2004) .

\bibitem{Brassard-review}  G. Brassard, A. Broadbent, A. Tapp, \emph{Quantum Pseudo-Telepathy}, Foundations of Physics, Volume 35, Issue 11, 1877-1907 (2005).

\bibitem{BMMN} M. Braverman, K. Makarychev, Y. Makarychev, A. Naor, \emph{The Grothendieck constant is strictly smaller than Krivine's bound}, Forum of Mathematics, Pi, Volume 1 (2013).

\bibitem{BBLV} J. Briët, H. Buhrman, T. Lee, T. Vidick,  \emph{Multipartite entanglement in XOR games. Quantum Information Processing} 13, 334-360 (2013).

\bibitem{BrVi} J. Briet, T. Vidick, \emph{Explicit lower and upper bounds on the entangled value of multiplayer XOR games}, Comm. Math. Phys. 321(1), (2013). 

\bibitem{BCPSW} N. Brunner, D. Cavalcanti, S. Pironio, V. Scarani, S. Wehner, \emph{Bell nonlocality}, Rev. Mod. Phys. 86, 419 (2014).

\bibitem{BCMW} H. Buhrman, R. Cleve, S. Massar, R. de Wolf, \emph{Non-locality and Communication Complexity}, Rev. Mod. Phys. 82, 665 (2010).

\bibitem{BRSW} H. Buhrman, O. Regev, G. Scarpa, R. de Wolf,  \emph{Near-Optimal and Explicit Bell Inequality Violations},  IEEE Conference on Computational Complexity 2011: 157-166.

\bibitem{Carne} T. K. Carne, \emph{Banach Lattices and Extensions of Grothendieck?s Inequality}, J. London Math. Soc., 21 (3), 496-516 (1980).

\bibitem{CHSH} J.F. Clauser, M.A. Horne, A. Shimony, R. A Holt, \emph{Proposed Experiment toTest Local-Hidden-Variable
Theories}, Phys. Rev. Lett. 23, 880 (1969).

\bibitem{CGLMP} D. Collins, N. Gisin, N. Linden, S. Massar, S. Popescu, \emph{Bell inequalities for arbitrarily high dimensional systems}, Phys. Rev. Lett. 88(4), 040404 (2002).

\bibitem{DeFl} A. Defant and K. Floret, \emph{Tensor Norms and Operator Ideals}, North-Holland, (1993).

\bibitem{DrOl} R.C. Drumond, R.I. Oliveira, \emph{Small violations of full correlations Bell inequalities for multipartite pure random states}, Physical Review A, 86 (1), 012117 (2012).

\bibitem{EPR} A. Einstein, B. Podolsky, N. Rosen, \emph{Can Quantum-Mechanical Description of Physical Reality Be Considered Complete?}, Phys. Rev. 47, 777 (1935).

\bibitem{GJPV}  C. E. Gonz\'alez-Guill\'en, C. H. Jim\'enez, C. Palazuelos, I. Villanueva, \emph{Sampling quantum nonlocal correlations with high probability}. Available in arXiv:1412.4010.

\bibitem {GPV} C. Gonz\'alez-Guill\'en, C. Palazuelos, I. Villanueva, \emph{Distance between Haar unitary and random gaussian matrices}. Available in : arXiv:1412.3743.
 
\bibitem{JP} M. Junge, C. Palazuelos, \emph{Large violation of Bell inequalities with low entanglement}, Comm. Math. Phys. 306 (3), 695-746 (2011).

\bibitem{JPPVWI} M. Junge, C. Palazuelos, D. P\'erez-Garc\'ia, I. Villanueva, M.M. Wolf, \emph{Unbounded violations of bipartite Bell Inequalities via Operator Space theory}.  Comm. Math. Phys. 300 (3), 715-739 (2010).

\bibitem{JPPVWII} M. Junge, C. Palazuelos, D. P\'erez-Garc\'ia, I. Villanueva, M.M. Wolf, \emph{Operator Space theory: a natural framework for Bell inequalities}, Phys. Rev. Lett. 104, 170405 (2010).

\bibitem{KRT} J. Kempe, O. Regev, B. Toner, \emph{The Unique Games Conjecture with Entangled Provers is False}, Proceedings of 49th Annual IEEE Symposium on Foundations of Computer Science (FOCS 2008), quant-ph/0710.0655 (2007).

\bibitem{LHBR} Y.-C. Liang, N. Harrigan, S. D. Bartlett, T. Rudolph, \emph{Nonclassical Correlations from Randomly Chosen Local Measurements}, Phys. Rev. Lett. 104, 050401 (2010).

\bibitem{LVB} Y.-C. Liang, T. Vertesi, N. Brunner, \emph{Device-independent bounds on entanglement}. Available in  arXiv:1012.1513.

\bibitem {MaPa} V. A. Marcenko, L. A. Pastur, \emph{Distribution of eigenvalues for some sets of random matrices}, Math. USSR Sbornik, 1:457-483, (1967).

\bibitem{Mermin} D. Mermin, \emph{Extreme quantum entanglement in a superposition of macroscopically distinct states}, Phys Rev Lett. 65 (15), 1838-1840 (1990).

\bibitem{KhVi} S. Khot, N. Vishnoi, \emph{The unique games conjecture, integrality gap for cut problems and embeddability of negative type metrics into $\ell_1$}, In Proceedings of 46th IEEE FOCS, pages 53.62 (2005).

\bibitem{Palazuelos II} C. Palazuelos, \emph{On the largest Bell violation attainable by a quantum state}, J. Funct. Anal., 267 (7), 1959-1985 (2014).

\bibitem{PWPVJ} D. P\'{e}rez-Garc\'{\i}a, M.M. Wolf, C. Palazuelos, I. Villanueva, M. Junge, \emph{Unbounded violation of tripartite Bell inequalities}, Comm. Math. Phys. 279 (2), 455-486 (2008).

\bibitem{Pisier I} G. Pisier, \emph{Grothendieck's Theorem, past and present}, Bull. Amer. Math. Soc. 49, 237-323 (2012). See also  arXiv:1101.4195.

\bibitem{Pisier II} G. Pisier, \emph{Tripartite Bell inequality, random matrices and trilinear forms}. Available in arXiv:1203.2509.

\bibitem{Pitowsky} I. Pitowsky, \emph{Macroscopic objects in quantum mechanics: A combinatorial approach}, Phys. Rev. A 70, 022103-1-6 (2004).

\bibitem{Raz} R. Raz, \emph{A Parallel Repetition Theorem}, SIAM Journal on Computing 27, 763-803 (1998).

\bibitem{Regev} O. Regev, \emph{Bell Violations through Independent Bases Games},  Quantum Inf. Comput., 12(1-2): 9-20 (2012).

\bibitem{SaZy} R. Salem, A. Zygmund, \emph{Some properties of trigonometric series whose terms have random signs}, Acta Mathematica 91(1), 245-301 (1954).

\bibitem{SVLBO}  P. Shadbolt, T. Vertesi, Y.-C. Liang, C. Branciard, N. Brunner, J. O'Brien, \emph{Guaranteed violation of a Bell inequality without aligned reference frames or calibrated devices}, Scientific Reports 2, 470 (2012).

\bibitem{SWZ} S. Szarek, E. Werner, K. Zyczkowski, \emph{How often is a random quantum state $k$-entangled?},  J. Phys. A: Math. Theor. 44, 045303 (2011). 

\bibitem{Tonge}  A. Tonge, \emph{The Von Neumann inequality for polynomials in several Hilbert-Schmidt operators},
J. London Math. (2) 1, 519-526 (1978).

\bibitem{Tsirelson} B. S. Tsirelson, \emph{Some results and problems on quantum Bell-type inequalities}, Hadronic J. Supp. 8(4), 329-345 (1993).

\bibitem{ViWe} T. Vidick, S. Wehner, \emph{More nonlocality with less entanglement}, Phys. Rev. A 83, 052310 (2011).

\bibitem{WLB}  J. J. Wallman, Y.-C. Liang, S. D. Bartlett, \emph{Generating nonclassical correlations without fully aligning measurements}, Phys. Rev. A 83, 022110 (2011).

\bibitem{WeWo I}  R.F. Werner, M.M. Wolf, \emph{Bell inequalities and Entanglement}, Quant. Inf. Comp. 1(3), 1-25 (2001).

\bibitem{WeWo II} R.F. Werner, M.M. Wolf, \emph{All multipartite Bell correlation inequalities for two dichotomic observables per site}, Phys. Rev. A 64, 032112 (2001).

\bibitem{ZuBr} M. Zukowski, C. Brukner, \emph{Bell's theorem for general $N$-qubit states}, Phys. Rev. Lett. 88, 210401 (2002).

\bibitem{ZPNC} K. Zyczkowski, K. A. Penson, I. Nechita, B. Collins, \emph{Generating random density matrices}, J. Math. Phys. 52 (6), 062201 (2011).

\end{thebibliography}
\end{document}